\documentclass[a4paper,11pt]{article}
\usepackage{jinstpub} 
\usepackage{lineno}
\usepackage{subfig}

\newcommand{\mX}{{M_X}}       
\newcommand{\sigmaX}{{\sigma}}

\newcommand{\mycomment}[1]{}
\newcommand{\met}{$E^{miss}$}  
\usepackage{float}

\notoctrue

\title{ADFilter - A Web Tool for New Physics Searches With Autoencoder-Based Anomaly Detection Using Deep Unsupervised Neural Networks}

\author[a]{S.~Chekanov,}
\author[b]{W.~Islam,}
\author[b]{R.~Zhang}
\author[a]{and N.~Luongo}

\affiliation[a]{HEP Division, Argonne National Laboratory, USA}
\affiliation[b]{Department of Physics, University of Wisconsin, Madison, WI, USA.}

\emailAdd{chekanov@anl.gov, wasikul.islam@cern.ch, rui.zhang@cern.ch, nluongo@anl.gov}

\abstract{
A web-based tool called ADFilter was developed to process collision events using autoencoders based on a deep unsupervised neural network. The autoencoders are trained on a small fraction of either collision data or Standard Model Monte Carlo simulations. The tool calculates loss distributions for input events, helping to determine the degree to which the events can be considered anomalous. It also calculates two-body invariant masses both before and after the autoencoders, as well as cross sections. 
Real-life examples are provided to demonstrate how the tool can be used to reinterpret existing LHC results with the goal of significantly improving exclusion limits.
}


\arxivnumber{ } 

\note{Preprint: ANL-HEP-190964, Sep 3, 2024}

\begin{document}

\maketitle
\flushbottom

\clearpage


\section{Introduction}
\label{sec:intro}

Traditionally, searches for new physics involve applying selection cuts to isolate events where sensitivity to Beyond Standard Model (BSM) phenomena is highest. These cuts are typically designed by exploring the kinematics of BSM events or using some general principles of how new physics might modify events produced by Standard Model (SM) processes.

Alternatively, selections can be performed by creating neural networks that "remember" the primary kinematic characteristics of data, which are predominantly governed by SM processes. These events are then removed, leaving only those that cannot be fully reconstructed by such networks. In this case, no prior knowledge of BSM models is required, making this selection method more agnostic to potential new physics. An overview of anomaly detection methods is provided in \cite{BELIS2024100091}.

The first LHC article utilizing anomaly detection to identify LHC events with five types of reconstructed objects (jets, $b$-jets, electrons, muons and photons) that are most likely to contain BSM scenarios was recently published by ATLAS \cite{ATLAS:2023ixc}. In this approach, autoencoders (AEs) are trained on a small fraction of the data to determine the global characteristics of event kinematics. The trained AE is then applied to the rest of the data that need to be tested. By analyzing the loss distributions, or the errors with which the AE reproduces the data, one can select only events with large AE losses. This process identifies outlier events that deviate significantly from the bulk of events, which are dominated by SM processes. The neural network approach is expected to capture full correlations between various inputs, making it superior to methods that rely purely on statistical measures to determine how unique an event is. Once an "anomalous" region of phase space with large loss values for events is determined, it must be further investigated using various techniques. The ATLAS study \cite{ATLAS:2023ixc} used searches in dijet invariant masses in the anomaly region to detect BSM signals. Additional demonstrations of this method using Monte Carlo (MC) simulations are provided in \cite{Chekanov:2021pus,Chekanov:2023uot}.

From an experimental standpoint, the advantages of such approaches are clear: These model-agnostic selections lead to an improved sensitivity to BSM phenomena. However, if BSM signals are not found, experimental limits should be calculated for events in the outlier region. Such limits should include acceptance corrections, which are difficult to estimate outside the LHC experiments. 

Thus, to facilitate comparisons of data in the outlier region with proposed BSM models, a simple public tool is needed that can propagate event records from any arbitrary BSM model and report the acceptances in the anomaly regions as defined by experiments. This approach to anomaly detection requires the wide public availability of trained autoencoders. These AEs, trained on a portion of data (or SM MC simulations), will represent the kinematics of data (or SM processes) and can thus be used to calculate acceptance for any BSM model.

This paper introduces a new online tool called "ADFilter" \cite{adfilter}. The purpose of this tool is to process event records from collision events (such as those generated by Monte Carlo simulations) through AEs used in physics research. It calculates the AE loss distribution for collision events,
and then it applies a selection cut, accepting only events within the anomaly region. This functionality enables the evaluation of acceptance corrections, which are calculated as the ratio of events that pass the AE cut to the total number of input events. Therefore, ADFilter is crucial for comparing BSM scenarios with data in anomaly regions.


\section{Description of ADFilter}
\label{sec:description}

\begin{figure}[h] 
\begin{center} \includegraphics[width=0.85\textwidth]{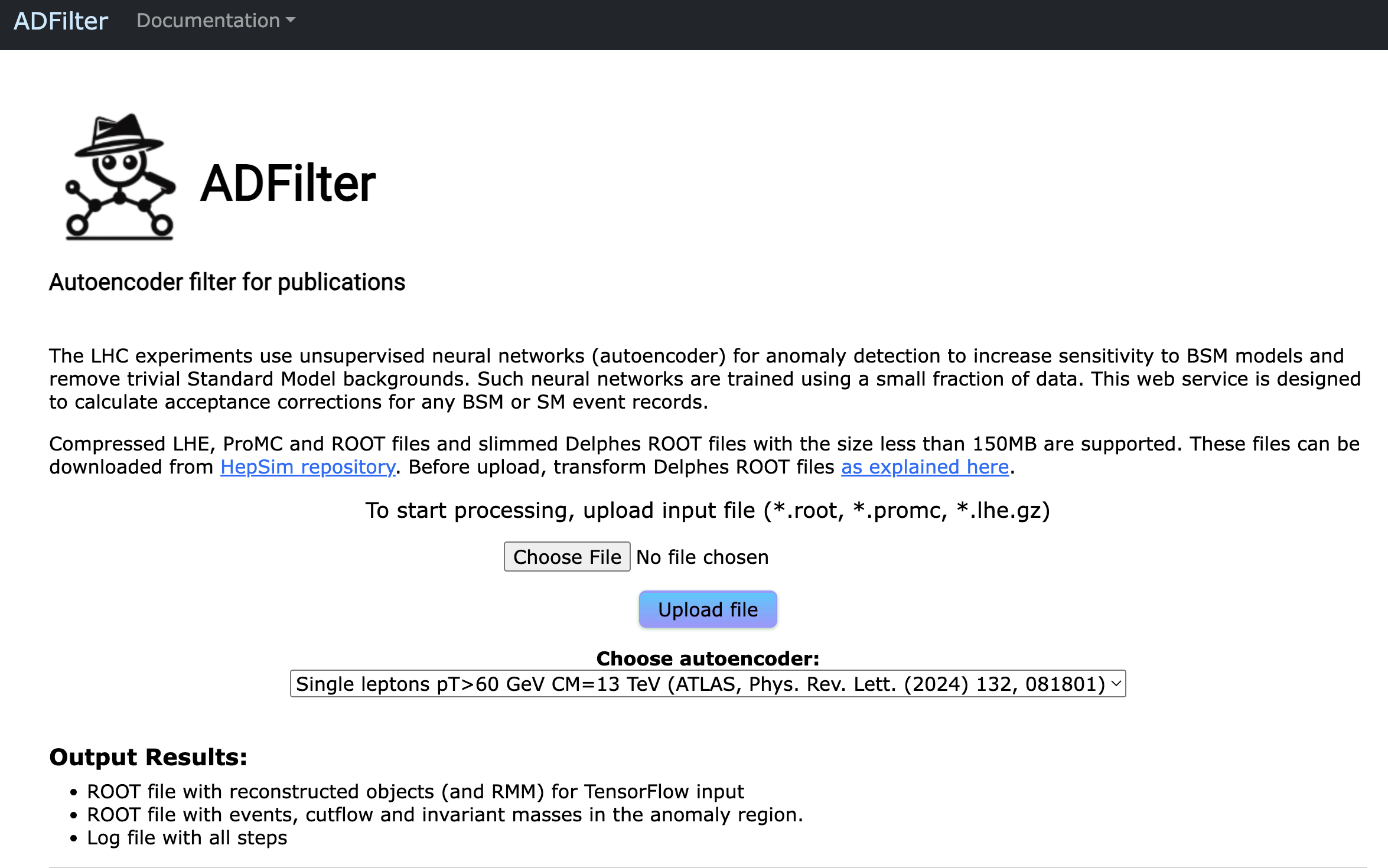} 
\end{center} 
\caption{A view of the web user interface for submission of input event records to the ADFilter.} 
\label{userGUI} 
\end{figure}

The ADFilter webpage \cite{adfilter} features the user interface shown in Fig.~\ref{userGUI}. The public version includes the AEs described in \cite{ATLAS:2023ixc} and \cite{Chekanov:2023uot}. The LHC paper \cite{ATLAS:2023ixc} employs a subset of the Run 2 ATLAS data, while the second paper relies on SM MC simulations for AE training. The autoencoder uses selection criteria based on single-lepton triggers, accepting events with at least one lepton with transverse momentum $p_T^l > 60$~GeV. This cut effectively reduces the multijet QCD background, which dominates the LHC data.

To start event processing, one needs to select the AE using a drop-down menu, then upload a file with input events. The formats for the input files are discussed in Sect.~\ref{sec:implementation}. The maximum size of the input file is 150 MB, which is typically sufficient for about 100,000 events in the file format to be discussed later.

After uploading, the ADFilter interface will submit several jobs in order to create:

\begin{itemize}

\item A ROOT file with the inputs for AE and a few test histograms. This ROOT file's name ends with the string "rmm.root". It contains basic kinematic distributions, the histogram "cross" with the observed cross section (in pb), the rapidity-mass matrix (RMM) \cite{Chekanov:2018nuh} for the first 50 events (for demonstration purposes), and the ROOT tree "inputNN", which stores non-zero values for the RMM and their indices.

\item A ROOT file with the final result. The name of this ROOT file contains the substring "ADFilter". It includes a histogram called "Loss", representing the numerical value of the reconstruction loss after processing through the AE. This histogram shows the success of the encoder-decoder process in reconstructing the compressed input. The "EventFlow" histogram shows the number of events entering the AE and the number of output events that exceed the "LossCut" value,  which is typically defined in the relevant publications. The output ROOT file also includes a set of histograms showing invariant masses before and after the AE, as well as the cross section in the selected anomalous region.

\item A text file that contains information about all processing steps. This file has the extension ".log". It can be used to monitor and verify each step of data processing, from the file with the input variables to the final file with the loss distribution. It also prints the selection cuts used for event processing.

\end{itemize}

All ROOT histograms can be viewed using the online browser and can also be downloaded for offline analysis.

It should be noted that if the input file contains only partons, this tool also applies parton showering and hadronization using the Pythia8 \cite{Sjostrand:2007gs} generator. This creates an intermediate binary file with the extension "promc", which includes all final-state particles. All technical details on how to use ADFilter are given in the menu called "Documentation."

The web interface of ADFilter is implemented in the PHP language with an external call to a binary program that converts the input data to ROOT files containing the tree with the input for the AE. The autoencoder calls were implemented in Python3. The jsROOT package \cite{Bellenot_2015} was used to display the output ROOT files online.


\section{Technical details on input files and event processing}
\label{sec:implementation}

The ADFilter web tool accepts a variety of input files containing collision events. These input files must be generated with the same center-of-mass collision energy as the events used for AE training.

The preferred event record should include the four-momentum of jets (including light-flavor jets and jets associated with $b$-quarks), electrons, muons, photons, and missing transverse energy. The jet definition should match the one used in the AE. An example of the generic input data structure is provided in Appendix~\ref{app}. The data format is a simple TTree called "Ntuple." An additional histogram ("meta") stores metadata with the center-of-mass energy (in GeV) used, whereas another histogram ("cross") contains the generator-level cross section (in pb) and the luminosity (in pb$^{-1}$) for the current run.

There are other types of input files that do not include reconstructed jets and leptons. These inputs can either be events with the four-momentum of final-state particles or files containing parton-level information. In the latter case, as previously mentioned, ADFilter propagates the events through the parton shower and hadronization. Then it applies jet reconstruction, isolation criteria for leptons, and the necessary transverse-momentum cuts as specified for the selected AE.

Below, we briefly describe the input files that can be used by ADFilter.

\subsection{Delphes input files}

The tool accepts ROOT ntuples generated by the Delphes fast simulation \cite{deFavereau:2013fsa}. However, because the input files can be quite large, Delphes files must be reduced in size ("slimmed"). To facilitate this, a simple tool is provided, as detailed on the "Documentation" page. This tool reduces the Delphes file to the smallest event structure that ADFilter can accept, as illustrated in Appendix~\ref{app}.

It is important to minimize selection cuts on Delphes objects to avoid introducing biases when using the AE. Generally, applying transverse momentum cuts of 30 GeV for all jets, leptons, and photons is the safest approach.

\subsection{Truth-level event record}
\label{promc}

To upload truth-level information with final-state particles, one can use ProMC files \cite{Chekanov:2013mma}. These files are based on Google's Protocol Buffers, which are a language-neutral, platform-neutral, and extensible mechanism for serializing structured data. They use "varints" as a way to store and compress integers using a variable number of bytes. Smaller numbers use fewer bytes. This means that low-energy particles (jets, clusters, cells, tracks, etc.) can be represented with fewer bytes since the values needed to represent such particles are smaller compared to those of high-energy particles or other objects.

ProMC files can be created directly from Monte Carlo generators or downloaded from the HepSim repository \cite{Chekanov:2014fga}, where about 350 event samples are stored in the ProMC format.

\subsection{LHE parton-level files}

One can also upload LHE files (Les Houches Event Files) \cite{Alwall:2006yp} directly from the Madgraph generator \cite{Alwall:2011uj, Alwall:2014hca}.
The files should be compressed and have the extension *.lhe.gz.
ADFilter applies the parton showering and hadronization. These steps are performed by the Pythia8 \cite{Sjostrand:2007gs} generator. The output events with final-state particles 
are stored in the ProMC file format, which is available for 
download.

\subsection{Object reconstruction step}

Collision events stored in binary ProMC files are processed with the goal of reconstructing jets, $b$-jets, isolated leptons, and photons. The jet definition and selection cuts are configured to match the objects used during AE training.

Jets, isolated electrons, and muons were reconstructed from stable particles. The jets were constructed with the anti-$k_T$ algorithm \cite{Cacciari:2008gp} as implemented in the {\sc FastJet} package~\cite{Cacciari:2011ma} with a distance parameter of $R=0.4$, which is commonly used in the ATLAS experiment. By default, the minimum transverse energy of all jets was $30$~GeV in the pseudorapidity range of $|\eta|<2.5$, but these values may change depending on the selection used by the AEs.

Leptons are required to be isolated using a cone of size $0.2$ in the azimuthal angle and pseudorapidity defined around the true direction of the lepton. All energies of particles inside this cone are summed. A lepton is considered isolated if it carries more than $90\%$ of the cone energy. The SM background processes require simulations of misidentification rates for muons and electrons ("fake rates"). We use a misidentification rate of 0.1\% for muons and 1\% for electrons. This is implemented by assigning a probability of $10^{-3}$ ($10^{-2}$) for a jet to be identified as a muon (electron) using a random number generator. The distributions were obtained for events having at least one isolated lepton with transverse momentum $p_T^{l}>30$~GeV and two jets with $p_T^{j}>30$~GeV.

\subsection{Pre-processing step}

The current version of ADFilter includes two autoencoder architectures, as previously mentioned, which require a pre-processing step to create input data for AEs. This step is described in this section. However, it's important to note that ADFilter is not limited to these two architectures; as a web tool, it can implement any autoencoder described in public research papers. 

After all objects (jets, leptons and photons) 
are available, a C++ program is used to transform kinematic features of each event
to the rapidity--mass matrix (RMM) which is proposed as an input for machine learning~\cite{Chekanov:2018nuh}. The RMM is a square matrix that includes reconstructed final states of jets, $b$-jets, muons, electrons, photons, and $E_{T}^{miss}$, where $E_{T}^{miss}$ is a single object, followed by $10$ ($b$-)jets and $5$ electrons, muons, and photons each, in descending order of transverse energy for each particle type.
This matrix contains characteristics of a single and a pair of all considered objects, filled with zeros to the matrix rows and columns where the corresponding objects are undetected.
In particular, the transverse energy of the leading objects or the transverse energy imbalances of the rest objects concerning the leading one in each object type are placed along the diagonal, where the transverse energy imbalance is defined as the ratio between their difference in the transverse energy and their sum.
The non-diagonal upper-right values of the matrix are the invariant mass of the corresponding two objects represented by the row and the column; in the case that $E_{T}^{miss}$ is involved, the transverse mass is used instead.
All values in the diagonal and above are scaled by $1/\sqrt{s}$ where $\sqrt{s}$ is the collision energy.
The non-diagonal lower-left values are $h_{ij} = C\cdot (\cosh((y_i-y_j)/2) - 1)$ where $y_i-y_j$ represents the rapidity difference between the two objects $i$ and $j$; when $j$ is \met, $h_{i} = C\cdot (\cosh(y_i) - 1)$ is used instead.
The constant $C$ was set to $0.15$~\cite{Chekanov:2018nuh}.

By construction, all elements of the RMM are defined to be between 0 and 1, and most variables are Lorentz-invariant under boosts along the longitudinal axis. 
To reduce biases in the shapes of the jet\;\!+\;\!$Y$ invariant mass spectra, the nine invariant mass variables are excluded from the RMM. The resulting input dimension is $36^2 - 9 = 1287$.
The RMM matrix is then flattened to a one-dimensional input vector before being fed into the AE.

\subsection{Autoencoder step}

To demonstrate ADFilter's capabilities, we will now describe the two autoencoders that are currently implemented.
After a ROOT file with the tree containing non-zero values of the RMM are created,
this tree was used as an input for the trained AE. The autoencoder is typically trained on a small
fraction of real collision events or the SM MC simulations, thus the weights of neurons represent  the snapshot of known physics events.
The training is not covered in this paper since
it is specific for each analysis domain. During the file upload, user should select the needed trained AE, description of which is usually given in the specific publication.

The AE is implemented using TensorFlow~\cite{abadi2016tensorflow}.
It comprises two sections, an encoder and a decoder.
The encoder compresses the input to a latent dimensional space, whereas the decoder takes the data in the latent layer and decompresses it back to its original size.
The network architecture for the encoder contains two hidden layers, with 800 and 400 neurons respectively, and a latent layer of 200 neurons.
The decoder reverses the structure of the encoder, using 400 and 800 neurons for the two hidden layers, and 1287 neurons for the output layer.
It compresses the inputs into lower dimensions using an encoder and then tries to reconstruct the original input using a decoder.
The AE architecture was optimized using tests with artificially created ``anomalous'' events by misidentifying objects in the Monte Carlo simulated events. The architecture reflects object multiplicities, particle identifications, and their kinematic characteristics well.
The Leaky ReLU~\cite{xu2015empirical} activation function is applied to the output in all hidden and output layers.

The final results of the AE step is to produce a ROOT file with invariant masses before and after applying a selection cut on the loss. The value of the selection cut is predetermined by the previous
publications.

The entire analysis chain, from the input file to the ROOT file with RMM and then the Python code with TensorFlow~\cite{abadi2016tensorflow}, was implemented using the BASH language. The website framework uses PHP with an external call to the master BASH script that executes the entire processing chain.

The public version of ADFilter uses the AE
described in \cite{ATLAS:2023ixc} trained on the Run2 LHC data with single-lepton pre-selections.  
In addition, the ADFilter menu includes
an AE trained on the SM events as explained in
\cite{Chekanov:2023uot}.

ADFilter can also be used as a command-line tool on the LXPLUS computing environment, as described in the menu "Documentation".


\section{Real-life examples}
\label{appHH}

The LHC publications focused on exclusion limits typically apply loose pre-selections guided by BSM model expectations. For example, one can apply a transverse momentum cut on leptons $p_T^l>60$~GeV 
in situations where high-momentum leptons are expected for certain BSM scenarios \cite{ATLAS2020}.
More sophisticated selections use boosted decision trees or neural networks trained on specific BSM events. While the first approach does not have the best sensitivity to BSM models, the second approach with improved sensitivity is significantly more time consuming to develop.

As discussed in the Introduction, another approach is to evaluate limits after applying AEs to reject the trivial SM background. In this approach, a single trained AE can be used for large variety of BSM scenarios, as long as event triggering is appropriate. To calculate limits after the AE, 
experimental data should be available in the anomaly region. Then, to set competitive exclusion limits, the BSM models need to be propagated through the same AE. 

This is where the ADfilter should be in hand: If the limits in the anomaly region
are available from LHC publications and the HEPdata database \cite{Maguire:2017ypu}, this tool helps 
estimate acceptance corrections for BSM events in order to compare the acceptance-corrected cross section with limits in anomaly regions. The easiest BSM models for testing are those
that create well-defined enhancements in two-body invariant masses. Because the tool reconstructs invariant masses
before and after cuts on the AE loss distribution, the acceptance can be easily estimated. 
In this section two examples of how to use ADFilter for two concrete BSM models are given.
This check normally takes 3-5 minutes using the ADFilter website. 

\subsection{Re-interpretation of sequential-standard model limits}

\begin{figure}[htb]
  \begin{center}
    \subfloat[Published LHC limits \cite{ATLAS2020}.]{\includegraphics[width=0.49\textwidth]{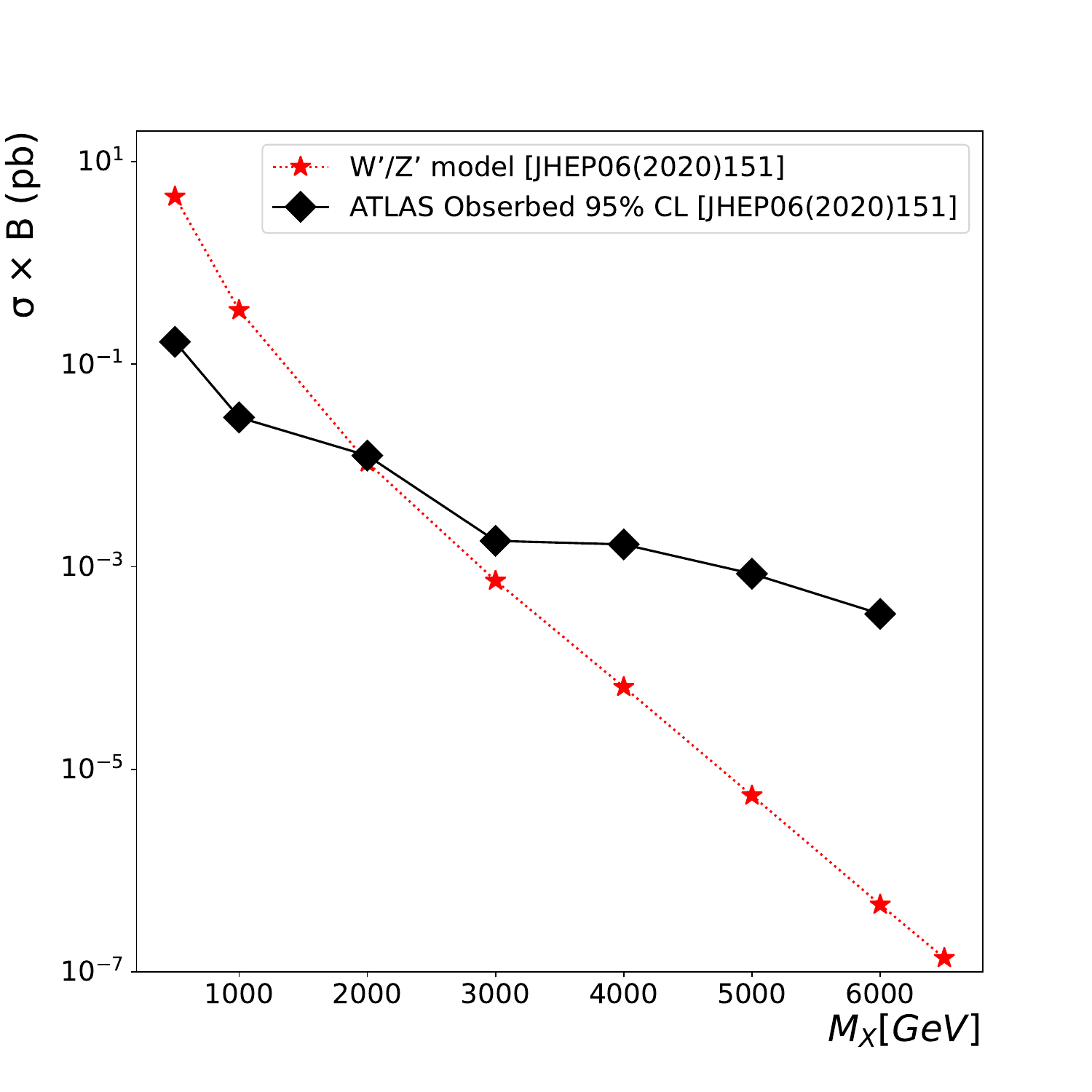}} 
    \subfloat[Re-interpreted limits using Ref.~\cite{ATLAS:2023ixc} and ADFilter.]{\includegraphics[width=0.49\textwidth]{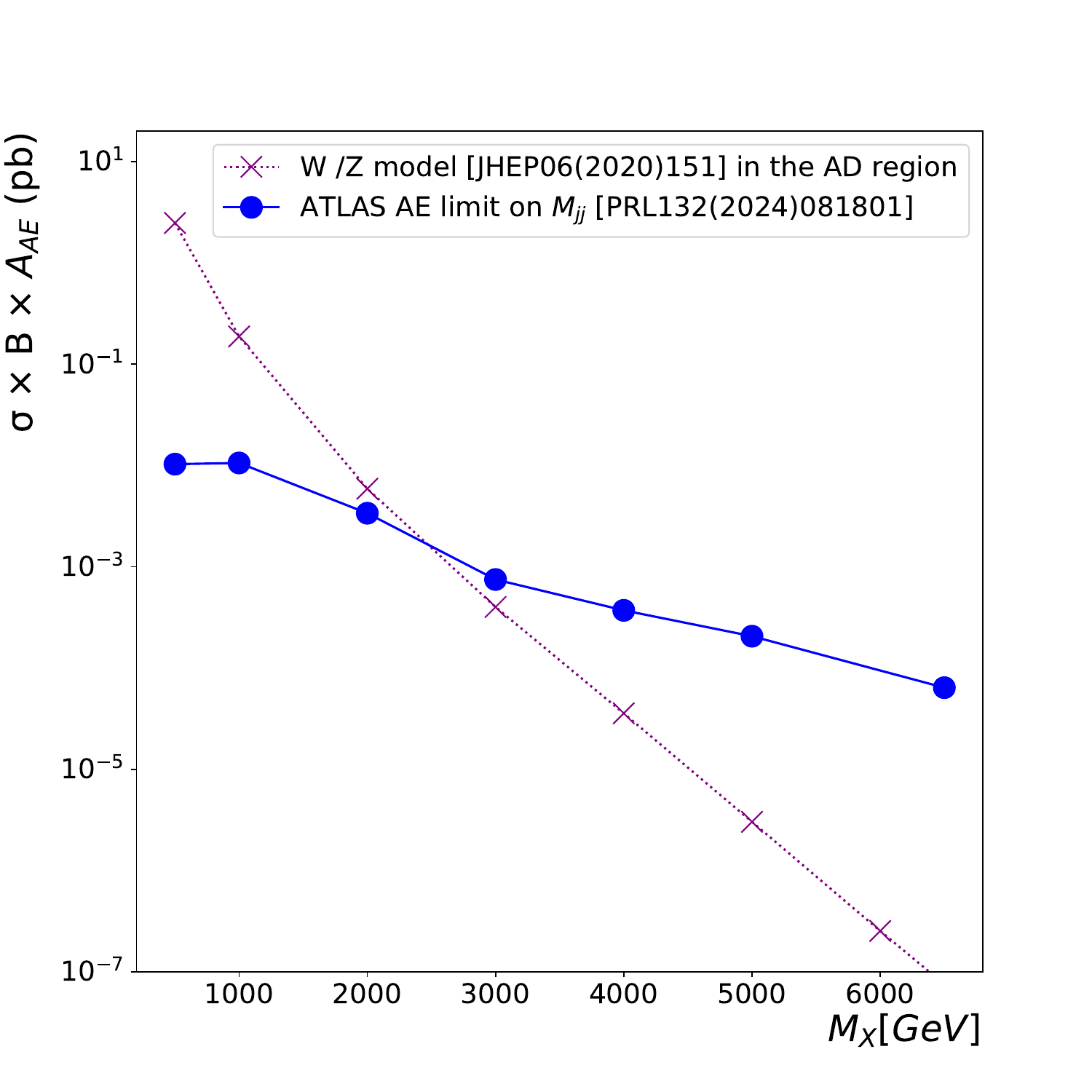}}
  \end{center}
\caption{
(a) Observed (black diamonds) 95\% credibility-level upper limit \cite{ATLAS2020} as a function of $Z'$ mass (denoted as $M_{X}$) on the cross-section ($\sigma$) times a branching ratio $B$ for the $W' \rightarrow Z' W^{\pm}$ production. The data selected with at least one isolated lepton with $p_{T}^l >60$~GeV. The red stars connected with dotted line show the SSM prediction.
(b) The filled circles connected by the blue line show
  the 95\% credibility-level upper limit with  the cut $p_{T}^l >60$~GeV plus the AE selection  \cite{ATLAS:2023ixc} for signals with
  the width of $\sigmaX/\mX=0.15$. 
  These limits are multiplied  by $(1/A_{sel} \times \varepsilon)$ from Ref.~\cite{ATLAS2020}. Only points for which this correction exists are shown.  Magenta crosses show the cross section in the anomaly region, i.e. after the additional AE correction obtained using the ADFilter tool.
}
\label{fig:SSM_2TeV}
\end{figure}

In the first example, we will utilize the publicly available data used to exclude the sequential-standard model (SSM) with the $W' \rightarrow Z' W^{\pm}$ process \cite{ATLAS2020}. 
The published 95\% CL upper limits are available as a function of masses $M_{X}$. For the SSM,  $X$ represents masses of $Z'$ extracted from the invariant mass of two jets, i.e. $M_{jj}$. Events were pre-selected by placing the transverse momentum cut $p_T^l>60$~GeV on leptons (electrons or muons). 
These limits are expressed in terms of $\sigma \times B$, where $\sigma$ is a cross section and $B$ is a branching ratio. The limits \cite{ATLAS2020} are reproduced as black diamonds in Fig.~\ref{fig:SSM_2TeV}(a). The red line shows the SSM model predictions. The original exclusion, reported in  \cite{ATLAS2020} was at 2 TeV, where black diamonds overlap with the red dashed line. 
Experimentally, such limits are corrected by the factors $A_{sel} \times \varepsilon$ as a function of the mass $\mX$, where $A_{sel}$  is the acceptance introduced by the event and object selection, and $\varepsilon$ is an object reconstruction efficiency. Such factors are known and available from publications.

The generic limits as a function of $\mX$ after applying the cut $p_T^l>60$~GeV and the AE are available \cite{ATLAS:2023ixc} and tabulated in HEPdata \cite{ATLAS:2023ixc}. These limits were calculated for the resonance width of $\sigmaX/\mX=0.15$, which is similar to the SSM studied in \cite{ATLAS2020}.  
These limits are obtained after the AE selection, therefore, they are expressed in terms of $\sigma  \times B \times \varepsilon \times A_{sel} \times  A_{AE}$, where $A_{sel}$ is the traditional acceptance introduced by event/object selections and
$A_{AE}$ is the acceptance introduced by the AE

The factor $A_{sel} \times \varepsilon$ is common to both \cite{ATLAS2020}  and \cite{ATLAS:2023ixc} publications, as long as selection cuts are the same. Therefore, one can multiply the limits \cite{ATLAS:2023ixc} by the correction factors $1/(A_{sel} \times \varepsilon)$ obtained from \cite{ATLAS2020}. 
After this correction, one obtains the  limits in terms of $\sigma \times B \times  A_{AE}$. 
There limits are shown in Fig.~\ref{fig:SSM_2TeV}(b) with the blue filled circles. They are substantially lower
than those expressed in terms of $\sigma \times B$, as the trained AE significantly reduces the rate of SM events (i.e. $A_{AE}<1$). 

In order to perform exclusions, the BSM model shown in Fig.~\ref{fig:SSM_2TeV}(a) needs to be re-interpreted in terms of $\sigma \times B \times  A_{AE}$. To do this, the ADFilter can be used to estimate the ADFilter acceptance $A_{AE}$.  This can be done by using a MC generator at truth level to produce $W' \rightarrow Z' W^{\pm}$ events for different mass points and uploading the generated LHE files to the ADFilter website. Our calculations estimate that the ADfilter acceptance is 55\% for SSM (on average). Then, the BSM prediction shown in \cite{ATLAS2020} should be multiplied by these acceptances as a function of $Z'$ mass.

The final result of re-interpretation of the limit is shown in  Fig.~\ref{fig:SSM_2TeV}(b). After anomaly detection, the mass exclusion moves from 2~TeV to 2.5~TeV,
where the blue filled circles overlap with the brown dotted line representing SSM corrected by the ADFilter acceptance.  This demonstrates that the application of the AE can increase the exclusion region by roughly 500 GeV.

It should be pointed out that our discussion assumes that the limits for Gaussian signals with the width 15\% are directly applicable to exclude the SSM. For high-precision studies, the availability of experimental limits for different realistic signal shapes in the anomaly region is highly desirable.

\subsection{Re-interpretation of charged Higgs $tbH^{+}$ limits}

\begin{figure}[htb]
  \begin{center}
    \subfloat[Published LHC limits \cite{ATLAS2020}.]{\includegraphics[width=0.49\textwidth]{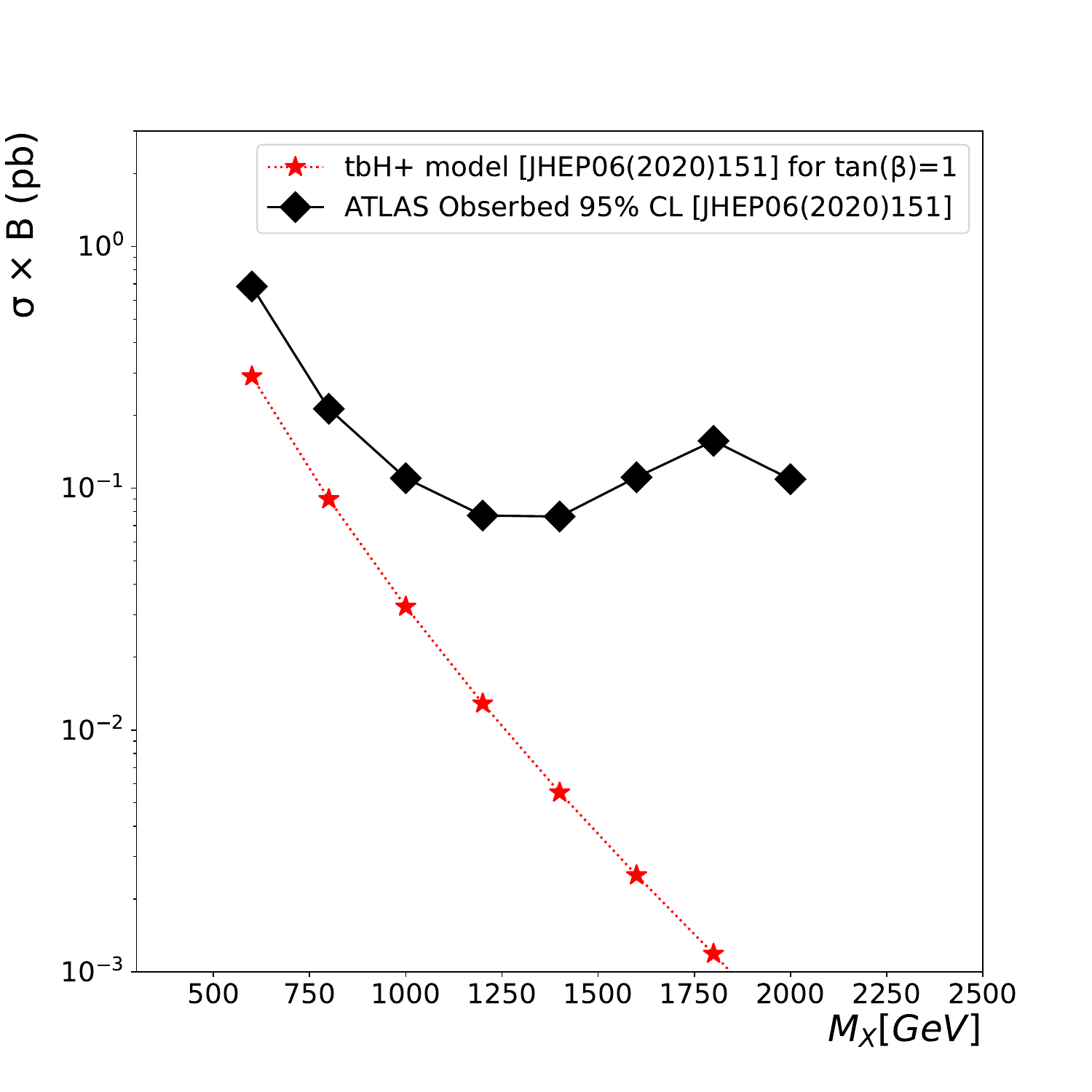}}
    \subfloat[Re-interpreted limits using Ref.\cite{ATLAS:2023ixc} and ADFilter.]{\includegraphics[width=0.49\textwidth]{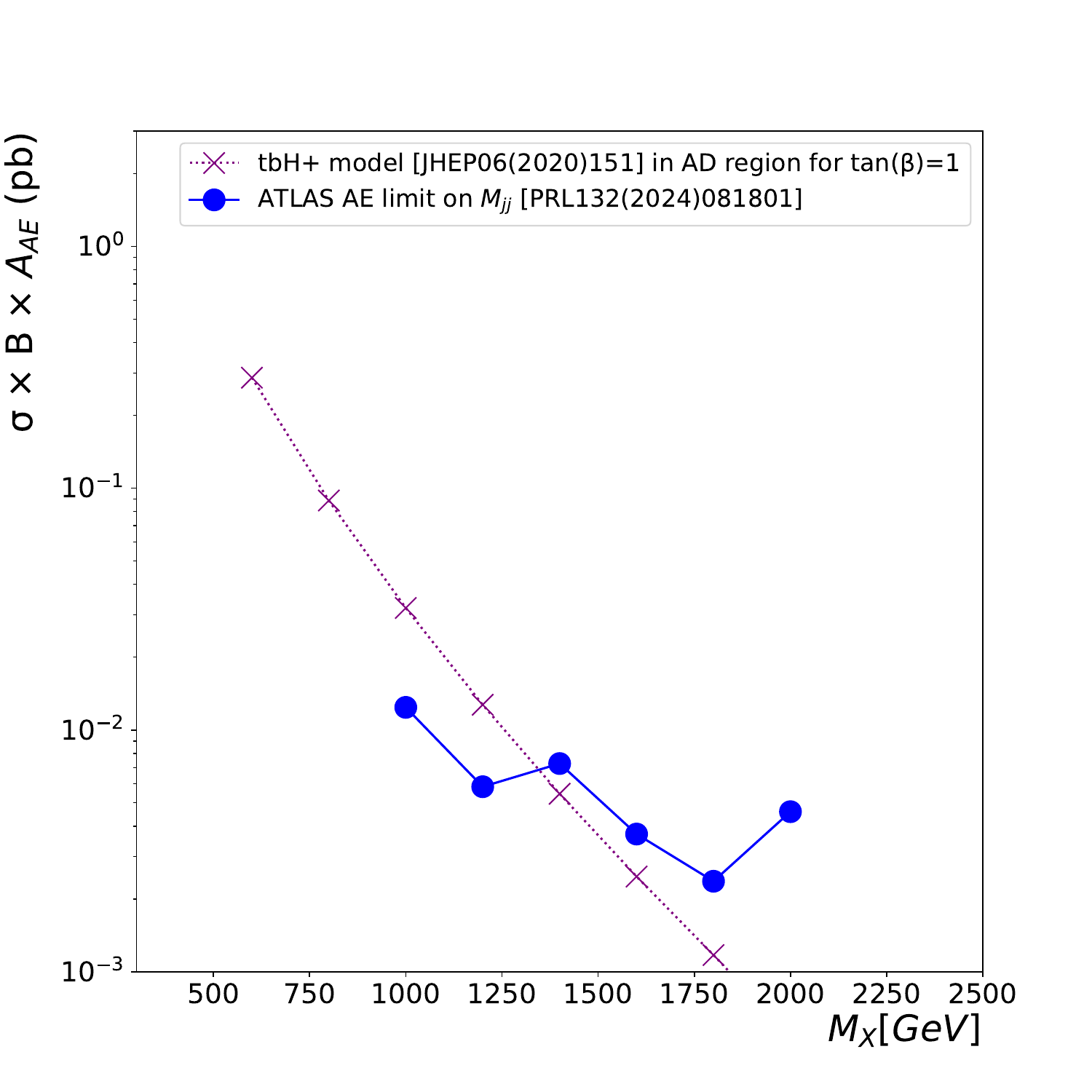}}
  \end{center}
 \caption{(a) Observed (filled diamonds) 95\% credibility-level upper limits  \cite{ATLAS2020} on  $\sigma\times B$ for the $tbH^{+}$ production cross section. The limit is shown as a function of $M_{X}$, where $X$ denotes the mass of $H^+$ derived from dijet masses. Events with at least one isolated lepton with $p_{T}^l >60$~GeV were used. Red filled stars show the  the  $tbH^{+}$ scenario with  $\tan (\beta)= 1$. (b) The 95\% credibility-level upper limits for Gaussian-shaped signals with the width of $\sigmaX/\mX=0.15$ published by the ATLAS collaboration, see Figure~4 of the publication \cite{ATLAS:2023ixc}. These limits are multiplied  by $(1/A_{sel} \times \varepsilon)$ from Ref.~\cite{ATLAS2020}. Only points for which this correction exists are shown. The magenta crosses show the observed cross section after the ADFilter correction, which is close to 95\%. Without anomaly detection, the $tbH^{+}$ process for $\tan (\beta ) = 1$ couldn't be excluded, but it is reliably excluded at 1.35~TeV after the AE. }
    \label{fig:Hplus_2TeV}
\end{figure}

Our second example is related to the charged Higgs $tbH^{+}$ process \cite{ATLAS2020}.
This process was modeled with MadGraph5\_aMC@NLO~\cite{Alwall:2014hca} at next-to-leading
order (NLO) in QCD~\cite{Degrande:2015vpa} based on a two-Higgs-doublet model (2HDM) using a four-flavour scheme implementation.
In this model,  $\tan (\beta)$ is the ratio of the two scalar doublets in the 2HDM.
Figure~\ref{fig:Hplus_2TeV}(a) reproduces the ATLAS observed limit (shown with black diamond symbols) for the pre-selection $p_T^l>60$~GeV as a function of the dijet mass and the $tbH^{+}$ process cross sections (red dotted line) for $\tan (\beta ) = 1$. 
The limits are shown as a function of $\mX$, where $X$ represents a $H^+$ mass derived from the mass of two jets, $M_{jj}$.
The $tbH^{+}$ process cross section does
not overlap with the observed limits, indicating that the data
do not have sufficient statistical power to exclude this BSM model with $\tan (\beta) = 1$.

The situation changes drastically when applying ADFilter. Figure~\ref{fig:Hplus_2TeV}(b)
shows the the limit with $p_{T}^l >60$~GeV plus the AE selection \cite{ATLAS:2023ixc} for signals with a width of $\sigmaX/\mX=0.15$. These limits were multiplied by the factor $1/(A_{sel} \times \varepsilon)$, reported in \cite{ATLAS2020}.  To justify this correction,
we assume that the width of $M_{jj}$ for $tbH^{+}$ process is roughly similar to
the width $\sigmaX/\mX=0.15$.
Then, the ADFilter was used to estimate the AE correction for the charged Higgs $tbH^{+}$ process created
with MadGraph5\_aMC@NLO. 
According to the calculated ADFilter acceptance, the observed cross section for the $tbH^{+}$ process does not change significantly after applying anomaly detection, since the reported ADFilter acceptance was above 95\%. 
The large AE acceptance indicates that the $tbH^{+}$ events are highly anomalous, compared to the SM events represented by the AE. 
Figure~\ref{fig:Hplus_2TeV}(b) shows the limit with  the cut $p_{T}^l >60$~GeV plus the AE selection  \cite{ATLAS:2023ixc} for signals with the width of $\sigmaX/\mX=0.15$. 
This creates a crossing point of the observed limits (shown with blue filled circles) at 1.35 TeV with the cross section from MadGraph5\_aMC@NLO after the 95\% correction by ADFilter.

Thus, even though the original ATLAS paper could not exclude $\tan (\beta) = 1$ scenario of the  $tbH^{+}$ process using the loose preselection with $p_{T}^l >60$~GeV, the exclusion becomes possible after applying ADFilter.
Note that the improvement in sensitivity is significantly greater than that for the SSM after the AE. The reason for this is due to the fact that the charged Higgs model is characterized by many additional objects (jets, $b-$jets), which makes such events significantly more complex and thus more "anomalous" than the SSM events.

It should be pointed out that if experimental limits are expressed in terms $\sigma \times B \times \varepsilon \times A_{sel} \times A_{AE}$ on generic signals after the AE, then a competitive exclusion of BSM models can be done by propagating truth-level event record via the ADFilter, which can estimate the factor $A_{sel} \times A_{AE}$ for BSM models. If BSM events were processed by the Delphes fast simulation, then the value $\varepsilon$ is also known.
Appendix ~\ref{app_lim} illustrate this approach using the limits for the Gaussian signal width of $\sigmaX/\mX=0.15$.
It compares the published limits (before and after the AE) with the BSM models after ADFilter.  These examples again demonstrate that the AE improves limits, or even makes the exclusion of BSM models possible. 

This discussion cannot be used to draw physics conclusions
on exclusion regions, since such questions require dedicated LHC studies using realistic
signal shapes for limits in the anomaly regions. Nevertheless, these examples confidently demonstrate the potential of ADFilter event selection and how this tool can be used by theorists. To facilitate such studies, broad public access to generic limits with various signal widths in the anomaly regions is highly anticipated.  

\section{Conclusion}
\label{sec:conclusion}

This paper introduces an online tool called ADFilter \cite{adfilter}. It is designed to process particle collision events using AEs trained on fractions of experimental data or SM Monte Carlo predictions. After processing the input events, the tool provides the loss values for these events and, optionally, the differential cross sections for two-body invariant masses. A higher loss value indicates that the events are more distinct in terms of kinematics compared to those used in training the autoencoder. 

By applying a selection cut on the loss value, the tool enables the calculation of acceptance values, which are then can be used to establish new exclusion limits based on currently available published limits.
ADFilter is particularly useful for quickly determining whether a given BSM scenario is sufficiently exotic, in terms of event kinematics, object multiplicity, object type, etc., compared to the bulk of LHC events used for autoencoder training. If the scenario yields a large loss value, it suggests that a generic selection based on the proposed approach could be applied to enhance sensitivity to such models, without the labor-intensive design of specific selection cuts. The tool reports the acceptance, which is crucial for comparing BSM models with limits provided by ATLAS publications using autoencoder-based selections.

A straightforward check that theorists can perform is as follows: If ADFilter reports a high acceptance rate (e.g., above 50\%) for a new BSM scenario implemented in LHE files from MadGraph, the corresponding events may be considered anomalous. This suggests a substantial opportunity within LHC studies to isolate and investigate these events further. If the LHC lacks dedicated studies for this scenario, ADFilter can effectively be used to rule out the BSM model, assuming that limits in the anomaly regions are published and accessible from HEPData \cite{Maguire:2017ypu}. Conversely, if a new proposed BSM model exhibits small loss values and low acceptance in the anomaly region, this would indicate that the model's features closely resemble those of SM kinematics. Consequently, detailed studies of such a model may require significant effort.

To enable effective exclusion of BSM models by the broad HEP community, public access to generic LHC limits with various signal widths in anomaly regions is highly desirable. This will allow the general HEP community to use ADFilter to exclude any arbitrary BSM scenario, provided that these models predict enhancements in invariant masses.

Finally, the tool is easily extendable to incorporate autoencoders from other publications.


\section{Acknowledgments}
The submitted manuscript has been created by UChicago Argonne, LLC, Operator of Argonne National Laboratory (“Argonne”). Argonne, a U.S. 
Department of Energy Office of Science laboratory, is operated under Contract No. DE-AC02-06CH11357. The U.S. Government retains for itself, 
and others acting on its behalf, a paid-up nonexclusive, irrevocable worldwide license in said article to reproduce, prepare derivative works, 
distribute copies to the public, and perform publicly and display publicly, by or on behalf of the Government.
The Department of Energy will provide public access to these results of federally sponsored research in accordance with the 
DOE Public Access Plan. \url{http://energy.gov/downloads/doe-public-access-plan}. Argonne National Laboratory’s work was 
funded by the U.S. Department of Energy, Office of High Energy Physics (DOE OHEP) under contract DE-AC02-06CH11357. The Askaryan Calorimeter Experiment was supported by the US DOE OHEP under Award Numbers DE-SC0009937, DE-SC0010504, and DE-AC02-76SF0051. WI and RZ are supported by DE-SC0017647. We gratefully acknowledge the computing resources provided by
the Laboratory Computing Resource Center at Argonne National Laboratory.

\newpage

\bibliographystyle{JHEP}
\bibliography{references}




\clearpage
\appendix
\part*{Appendix}
\addcontentsline{toc}{part}{Appendix}

\clearpage
\section{Example of the input data structure}
\label{app}
A simple ROOT TTree is used to store data used as the input for ADFilter. This is a simple PyROOT
script to create 1000 events with jets, b-jets, electrons, muons, photons and missing energy (MET).

The data are stored in the TTree called ``Ntuple".
The additional histogram "meta" keeps some metadata, such as the centre-of-mass energy (in GeV).

Run this example as ``python example.py", assuming that
``example.py" contains the lines of this code. This produces the ROOT file with one electron, one muon, one photon, one jet and b-jet.
All objects have transverse energy of 100 GeV.
When using the created file as input, the ADFilter reports $\log(Loss) = -9.822$ (see the histogram ``Loss" in the created file ``dummy\_root\_ADFilter.root").

\begin{verbatim}
#!/usr/bin/env python
#
# Create Dummy record for ADFilter
# python ./dummy.py --outputlist output.root  --cmsEnergy 13000 

# The necessary import(s):
import ROOT
import argparse
from array import array

## C/C++ style main function
def main( filename, cmsENERGY ):

    # The name of the application:
    APP_NAME = "dummy"
    NEvents=1000 # number of events
    
    import logging # Set up a logger object:
    logger = logging.getLogger( APP_NAME )
    logger.setLevel( logging.INFO )
    hdlr = logging.StreamHandler( sys.stdout )
    frmt = logging.Formatter( "%(name)-14s%(levelname)8s %(message)s" )
    hdlr.setFormatter( frmt )
    logger.addHandler( hdlr )

    #### ntuple writer ###
    cmsEnergy=cmsENERGY;
    outputFileName=filename
    outputFile=ROOT.TFile(outputFileName, "RECREATE");
    ntuple = ROOT.TTree("Ntuple", "Ntuple for ADFilter");

    N_JET =array('i', [0])
    JET_pt = ROOT.vector('Double32_t')()
    JET_eta = ROOT.vector('Double32_t')()
    JET_phi = ROOT.vector('Double32_t')()
    JET_mass = ROOT.vector('Double32_t')()

    N_bJET = array('i', [0])
    bJET_pt = ROOT.vector('Double32_t')()
    bJET_eta = ROOT.vector('Double32_t')()
    bJET_phi = ROOT.vector('Double32_t')()
    bJET_mass = ROOT.vector('Double32_t')()

    N_EL = array('i', [0])
    EL_pt = ROOT.vector('Double32_t')()
    EL_eta = ROOT.vector('Double32_t')()
    EL_phi = ROOT.vector('Double32_t')()

    N_MU = array('i', [0])
    MU_pt = ROOT.vector('Double32_t')()
    MU_eta = ROOT.vector('Double32_t')()
    MU_phi = ROOT.vector('Double32_t')()

    N_PH = array('i', [0])
    PH_pt = ROOT.vector('Double32_t')()
    PH_eta = ROOT.vector('Double32_t')()
    PH_phi = ROOT.vector('Double32_t')()
    PH_e = ROOT.vector('Double32_t')()

    N_MET = array('i', [0])
    MET_met = ROOT.vector('Double32_t')()
    MET_eta = ROOT.vector('Double32_t')()
    MET_phi = ROOT.vector('Double32_t')()

    Evt_Weight = ROOT.vector('Double32_t')()

    ntuple.Branch( 'JET_n', N_JET, 'N_JET/I' )
    ntuple.Branch( 'JET_pt', JET_pt, 256000, 0 )
    ntuple.Branch( 'JET_eta', JET_eta, 256000, 0 )
    ntuple.Branch( 'JET_phi', JET_phi, 256000, 0 )
    ntuple.Branch( 'JET_mass', JET_mass, 256000, 0 )

    ntuple.Branch( 'N_MET', N_MET, 'N_MET/I' )
    ntuple.Branch( 'MET_met', MET_met, 256000, 0 )
    ntuple.Branch( 'MET_eta', MET_eta, 256000, 0 )
    ntuple.Branch( 'MET_phi', MET_phi, 256000, 0 )

    ntuple.Branch( 'bJET_n', N_bJET, 'N_bJET/I' )
    ntuple.Branch( 'bJET_pt', bJET_pt, 256000, 0 )
    ntuple.Branch( 'bJET_eta', bJET_eta, 256000, 0 )
    ntuple.Branch( 'bJET_phi', bJET_phi, 256000, 0 )
    ntuple.Branch( 'bJET_mass', bJET_mass, 256000, 0 )

    ntuple.Branch( 'MU_n', N_MU, 'N_MU/I' )
    ntuple.Branch( 'MU_pt', MU_pt, 256000, 0 )
    ntuple.Branch( 'MU_eta', MU_eta, 256000, 0 )
    ntuple.Branch( 'MU_phi', MU_phi, 256000, 0 )

    ntuple.Branch( 'PH_n', N_PH, 'N_PH/I' )
    ntuple.Branch( 'PH_pt', PH_pt, 256000, 0 )
    ntuple.Branch( 'PH_eta', PH_eta, 256000, 0 )
    ntuple.Branch( 'PH_phi', PH_phi, 256000, 0 )
    ntuple.Branch( 'PH_e', PH_e, 256000, 0 )

    ntuple.Branch( 'EL_n', N_EL, 'N_EL/I' )
    ntuple.Branch( 'EL_pt', EL_pt, 256000, 0 )
    ntuple.Branch( 'EL_eta', EL_eta, 256000, 0 )
    ntuple.Branch( 'EL_phi', EL_phi, 256000, 0 )

    ntuple.Branch( 'Evt_Weight', Evt_Weight, 256000, 0 )

    meta = ROOT.TH1F("meta","meta",10,0,10);
    meta.Fill("CMS energy [GeV]",cmsEnergy);
    meta.Write();

    print("Total number of events=", NEvents)
    for entry in range( NEvents ):
        if (entry%10 == 0): logger.info( "Processing entry %i" % entry )
        # Print the properties of the electrons:

        JET_pt.clear();  JET_eta.clear(); JET_phi.clear(); JET_mass.clear();   
        bJET_pt.clear(); bJET_eta.clear(); bJET_phi.clear(); bJET_mass.clear();
        EL_pt.clear(); EL_eta.clear(); EL_phi.clear();
        MU_pt.clear(); MU_eta.clear(); MU_phi.clear();
        PH_pt.clear(); PH_eta.clear(); PH_phi.clear(); PH_e.clear();
        MET_eta.clear(); MET_phi.clear(); MET_met.clear();
        Evt_Weight.clear();

        # define electrons
        N_EL[0] = 1;
        for j in  range(N_EL[0]):
            EL_phi.push_back(0.0)
            EL_eta.push_back(0.0) 
            EL_pt.push_back(100.0)  # pt>100 GeV

        # define muons
        N_MU[0] =  1 
        for j in range(N_MU[0]):
            MU_phi.push_back(0.0)
            MU_eta.push_back(0.0)
            MU_pt.push_back(100.0)   # pt>100 GeV

        # define your photons
        N_PH[0] = 1 
        for j in range(N_PH[0]):  
            PH_phi.push_back(0.0)
            PH_eta.push_back(0.0)
            PH_pt.push_back(100.0)    # pt>100 GeV
            PH_e.push_back(0.0)

        # define here light-flavour jets
        N_JET[0] = 1
        for j in range(N_JET[0]):
            JET_phi.push_back(0)
            JET_eta.push_back(0.0)
            JET_pt.push_back(100.0)
            JET_mass.push_back(100.0)  # pt>100 GeV

        # define here b-jets
        N_bJET[0] = 1 
        for j in range(N_bJET[0]):git push   origin master;

            bJET_phi.push_back(0.0)
            bJET_eta.push_back(0.0)
            bJET_pt.push_back(100.0)
            bJET_mass.push_back(100.0)  # pt>100 GeV

        # define MET. 
        N_MET[0] = 1 # should be 1 always! 
        for j in range(N_MET[0]):
            MET_phi.push_back(0.0)
            MET_eta.push_back(0.0)
            MET_met.push_back(0.0)

        # define event weight
        Evt_Weight.push_back(1)
        Evt_Weight.push_back(1)

        ntuple.Fill()
           
    print("Write the TTree to the output file:",outputFileName);
    #outputFile.Write("",TFile.kOverwrite)
    ntuple.Write()
    #ntuple.Show(1)

    # Close the output file
    outputFile.Close();
    print("Write=",outputFileName)

    return 0

# Run the main() function:
if __name__ == "__main__":
   import sys

   parser = argparse.ArgumentParser(
       description="Extracts a few basic quantities \
       from the xAOD file and dumps them into a text file")
   parser.add_argument("--outputlist", help="Optional list of output ROOT files",
                       nargs='+', action="store", default=False)
   parser.add_argument("--cmsEnergy", help="Optional CMS energy",
                       nargs='+', action="store", default=False)

   args = parser.parse_args()

   cmsEnergy=13000
   filename="dummy.root"
   if args.outputlist:
            filename=args.outputlist[0] 
   if args.cmsEnergy:
            print("CMS energy =",args.cmsEnergy[0]);
            cmsEnergy=float(args.cmsEnergy[0]) 

   sys.exit( main( filename, cmsEnergy  ) )

\end{verbatim}

\clearpage
\section{Alternative representation of limits}
\label{app_lim}

This appendix discusses an alternative representation of BSM exclusions.  In this approach, one can use experimental limits obtained using Gaussian signal shapes without corrections by acceptance $A_{sel}$ and efficiency
$\varepsilon$. If a BSM model predicts signals in invariant mass distributions 
that are roughly similar to the Gaussian width,
then one can overlay the observed cross section of such models with the Gaussian limits.

\begin{figure}[htb]
  \begin{center}
    \subfloat[Published LHC limits \cite{ATLAS2020}.]{\includegraphics[width=0.49\textwidth]{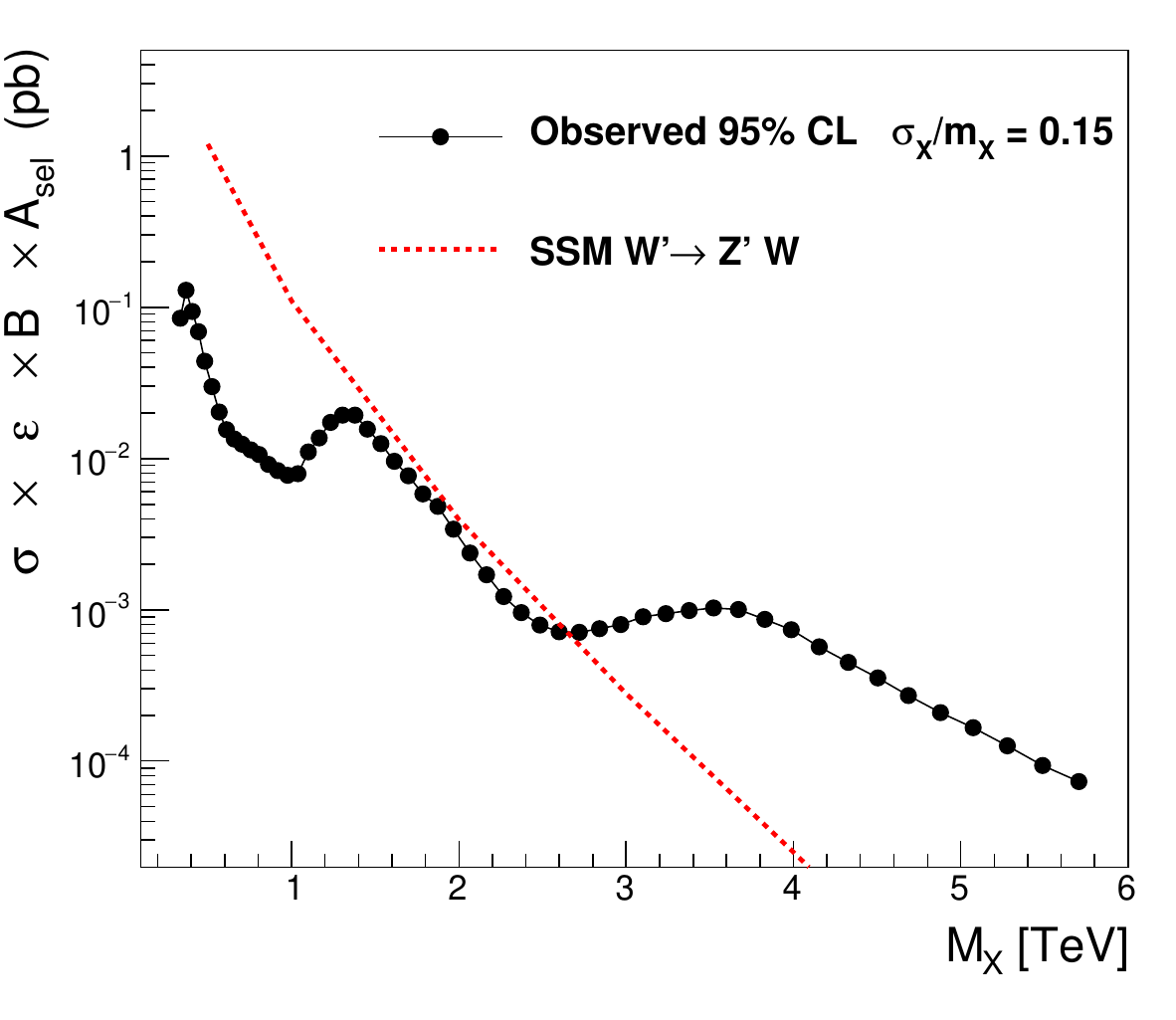}} 
    \subfloat[Re-interpreted limits using Ref.~\cite{ATLAS:2023ixc} and ADFilter.]{\includegraphics[width=0.49\textwidth]{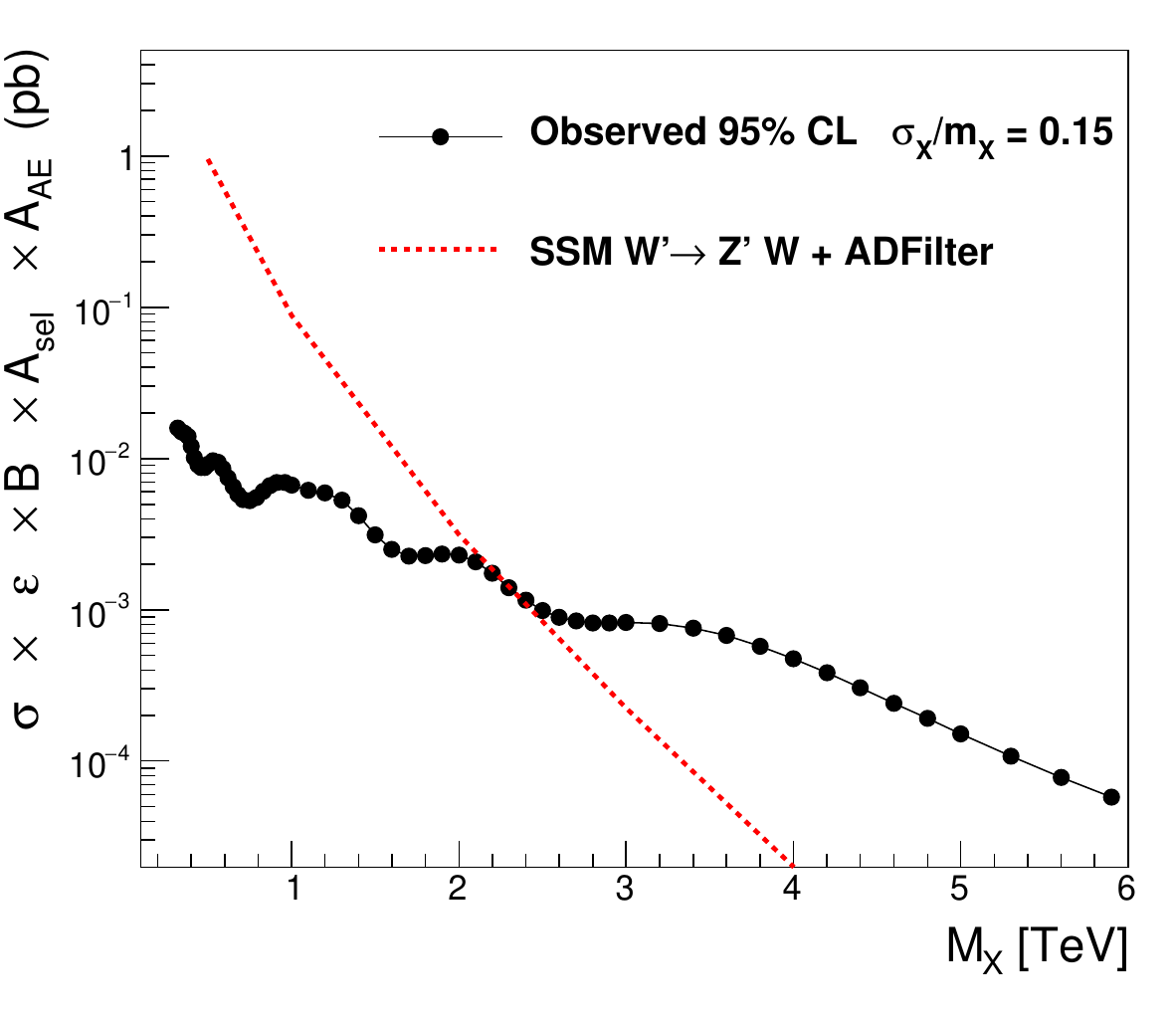}}
  \end{center}
\caption{
(a) Observed 95\% credibility-level upper limit \cite{ATLAS2020} as a function of $Z'$ mass (denoted as $M_{X}$) on the $\sigma \times \varepsilon \times B \times A_{sel}$ for  Gaussian signals with the width of $\sigmaX/\mX=0.15$. The events are selected with at least one isolated lepton with $p_{T}^l >60$~GeV. The red dotted line shows the SSM prediction.
(b) The 95\% credibility-level upper limit with  the cut $p_{T}^l >60$~GeV plus the AE selection for signals with the width of $\sigmaX/\mX=0.15$ \cite{ATLAS:2023ixc}. The theoretical prediction for the SSM is corrected by the ADFilter acceptance. 
}
\label{figa:SSM_2TeV}
\end{figure}

\begin{figure}[htb]
  \begin{center}
    \subfloat[Published LHC limits \cite{ATLAS2020}.]{\includegraphics[width=0.49\textwidth]{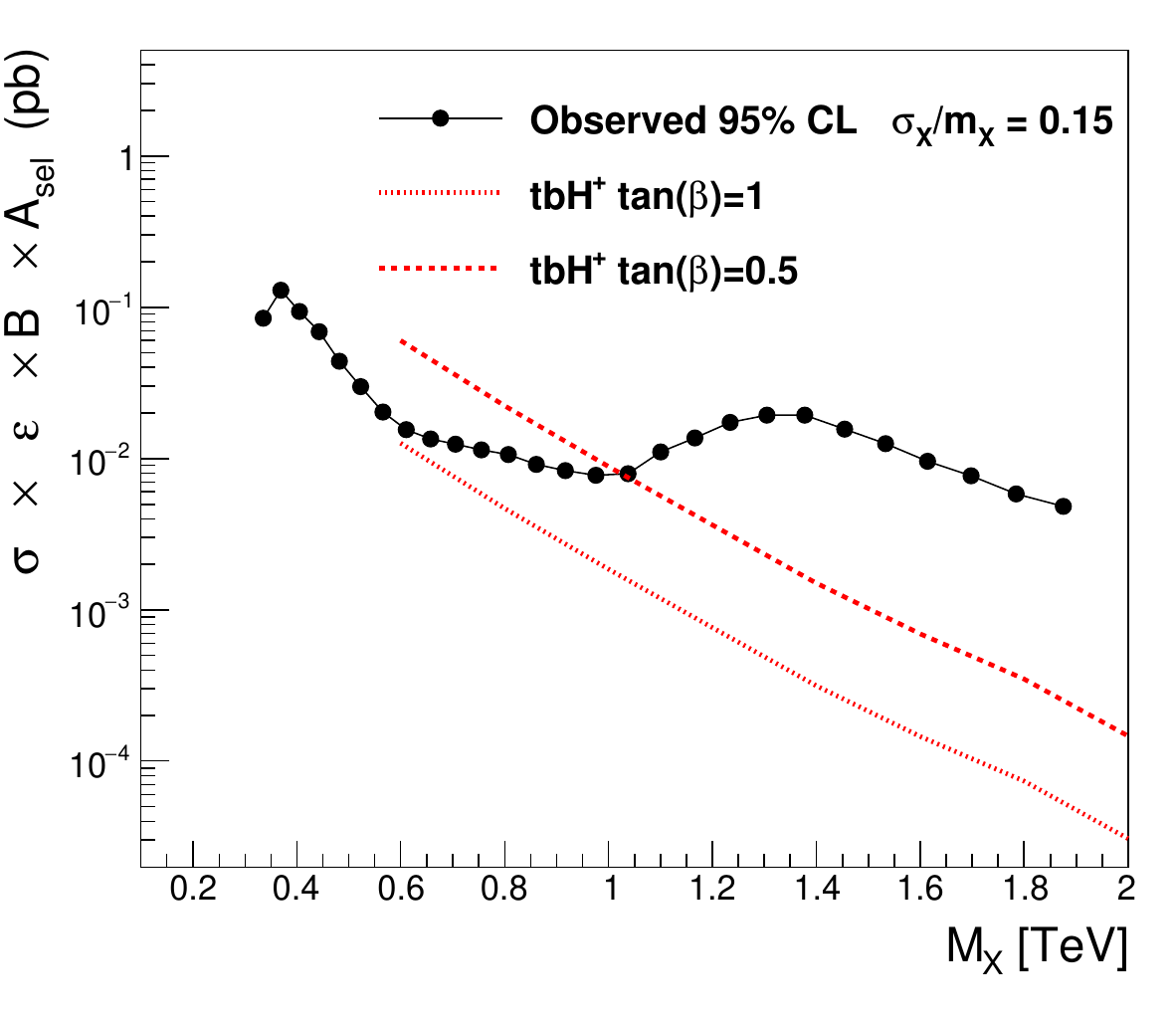}}
    \subfloat[Re-interpreted limits using Ref.\cite{ATLAS:2023ixc} and ADFilter.]{\includegraphics[width=0.49\textwidth]{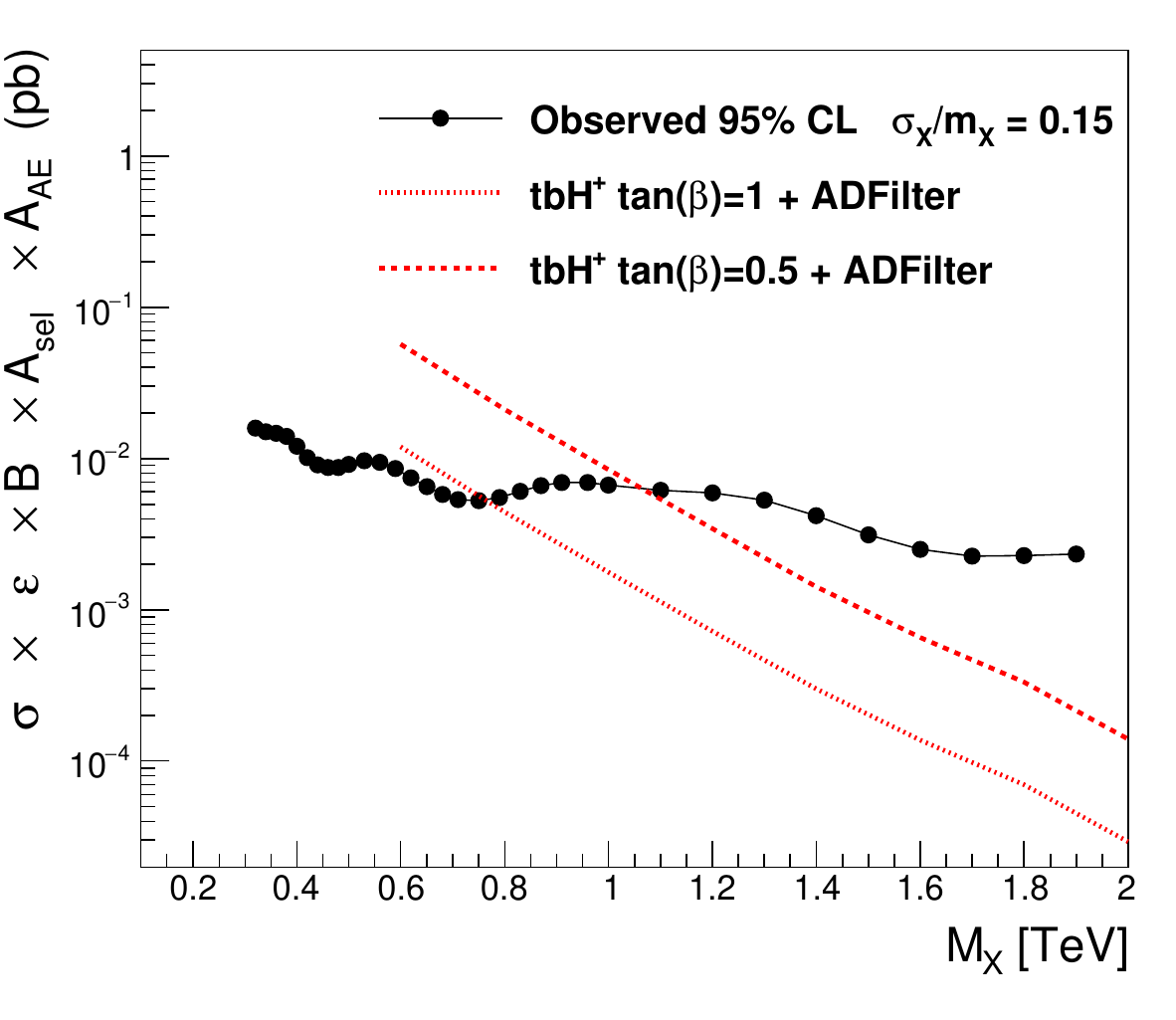}}
  \end{center}
 \caption{(a) Observed 95\% credibility-level upper limits  \cite{ATLAS2020} on  the $\sigma \times \varepsilon \times B \times A_{sel}$
for  Gaussian signals with the width of $\sigmaX/\mX=0.15$.  Events with at least one isolated lepton with $p_{T}^l >60$~GeV were used. Red dotted and dashed lines show the  the  $tbH^{+}$ scenarios with   $\tan (\beta) = 0.5$ and  $\tan (\beta) = 1$. (b) The 95\% credibility-level upper limits for Gaussian-shaped signals with the width of $\sigmaX/\mX=0.15$ published by the ATLAS collaboration, see Figure~4  of the publication \cite{ATLAS:2023ixc}.  Red lines show the observed cross section after the ADFilter correction, which is close to 95\%. Without anomaly detection, the $tbH^{+}$ process for $\tan (\beta) = 1$ couldn't be excluded, but with AE, this model is excluded. }
    \label{figa:Hplus_2TeV}
\end{figure}

Figures~\ref{figa:SSM_2TeV}  and \ref{figa:Hplus_2TeV} show such Gaussian limits. These figures also
demonstrate how the limits can be improved by applying the AE and the ADFilter for the signal BSM models.
For SSM, the limits confidently exclude the mass region below 2.5~TeV.
In the case of charged Higgs, the AE allows us to exclude $btH^+$ with $\tan (\beta) = 1$.
Note that the excluded mass point is somewhat lower than for Fig.~\ref{fig:Hplus_2TeV}(b), since the assumption on the Gaussian signal shape with the width $\sigmaX/\mX=0.15$ may not be appropriate for the $tbH^+$ model. 

Note that, typically, Gaussian limits can be calculated for fine steps in invariant masses, which is more difficult to achieve for BSM-specific limits based on MC event samples.
The limits shown in Fig.~\ref{fig:SSM_2TeV}(b) and \ref{fig:Hplus_2TeV}(b) use only a few points for which $A_{sel}$ and $\varepsilon$ exist \cite{ATLAS2020}, and the extrapolation between the points may introduce additional disrepancy with fine-binned limits shown in Fig.~\ref{figa:SSM_2TeV}  and \ref{figa:Hplus_2TeV}.

\end{document}